# Forecasting Avalanches in Branched Actomyosin Networks with Network Science and Machine Learning

Chengxuan Li[1,2], James Liman[2,3], Yossi Eliaz[1,2], and Margaret S. Cheung[1,2,3,4,*]

[1]*Department of Physics, University of Houston, Houston, Texas 77204, USA*

[2]*Center for Theoretical Biological Physics, Rice University, Houston, Texas 77005, USA*

[3]*Department of Bioengineering, Rice University, Houston, Texas 77030, USA*

[4]*Pacific Northwest National Laboratory, Seattle, WA, 98109, USA*

[*]Corresponding author: Margaret S. Cheung, Pacific Northwest National Laboratory, Seattle, Washington, margaret.cheung@pnnl.gov

**Abstract**

We explored the dynamic and structural effects of actin-related proteins 2/3 (Arp2/3) on actomyosin networks using mechanochemical simulations of active matter networks. On the nanoscale, the Arp2/3 complex alters the topology of actomyosin by nucleating a daughter filament at an angle with respect to a mother filament. At a subcellular scale, they orchestrate the formation of a branched actomyosin network. Using a coarse-grained approach, we sought to understand how an actomyosin network temporally and spatially reorganizes itself by varying the concentration of the Arp2/3 complexes. Driven by motor dynamics, the network stalls at a high concentration of Arp2/3 and contracts at a low Arp2/3 concentration. At an intermediate Arp2/3 concentration, however, the actomyosin network is formed by loosely connected clusters that may collapse suddenly when driven by motors. This physical phenomenon is called an "avalanche" largely due to the marginal instability inherent to the morphology of a branched actomyosin network when the Arp2/3 complex is present. While embracing the data science approaches, we unveiled the higher-order patterns in the branched actomyosin networks and discovered a sudden change in the "social" network topology of actomyosin, which is a new type of avalanches in addition to the two types of avalanches associated with a sudden change in the size or shape of the whole actomyosin network, as shown in a previous investigation. Our new finding promotes the importance of using network theory and machine learning models to forecast avalanches in actomyosin networks. The mechanisms of the Arp2/3 complexes in shaping the architecture of branched actomyosin networks obtained in this paper will help us better understand the emergent reorganization of the topology in dense actomyosin networks that are difficult to detect in experiments.

# I. Introduction

In muscle cells, actin filaments and myosins are organized into a striped sarcomere [1], and in non-muscle cells, actomyosin networks tend to be isotropic, especially at the edge of cells such as the actin cortex [1, 2]. We hypothesized that the nanostructure of the actomyosin network dictates the structure and dynamics of the entire system [1]. Actin-binding proteins (ABPs) are the key drivers of changes in the local structure of actomyosin networks. One of such ABP is the Arp2/3 complex [3] which is responsible for geometrical arrangement in a global architecture by being a nucleator for branched actomyosin networks [4, 5]. The Arp2/3 complex, also known as a brancher in the system, creates a junction of a daughter filament nucleated from its mother filament, and subsequently orchestrates the formation of branched actin networks [3]. Together with myosin motors, the branched actomyosin networks are responsive to mechanical perturbations from the environment of a cell [6], as the network organization controls contractile tension generation in a cell.

Computational models have been used to explore actomyosin contractility [7, 8], but few of these computational models explore the effect of the Arp2/3 complex on the dynamics of the system. Despite extensive experimental and computational studies [9-12], only recently has the computational work from our group shown that the presence of the Arp2/3 complex causes sudden collapse dynamics of marginally stable actomyosin networks, called "avalanches" [13]. The avalanche possibly underscores the phenomenon of a 'cytoquake' [14], a drastic structural change in the actomyosin network within a short period of time. The biophysical importance of this phenomenon is appreciated since it is deeply related to the structural rearrangement of the cytoskeleton. However, how this phenomenon connects to the nanoarchitecture of the branched actomyosin network orchestrated by the Arp2/3 complexes remains unclear.

In this work, we explored the impact of Arp2/3 concentration in modulating avalanches in branched actomyosin networks using coarse-grained simulations. We used the software package MEDYAN [15], which simulates the organization of actin filaments with mechanochemical feedback from actin-binding proteins, such as those that form catch bonds (non-muscle myosin IIA motors, NMIIA) [16-18], slip bonds (α-actinin linkers [18, 19]), and filament nucleators (Arp2/3 complexes [9, 13]). In our simulations, even though we only changed the concentration of Arp2/3 complexes while the turnover rates [1, 20] remained the same in the chemical reactions, the patterns of the temporally evolving networks were incredibly complex. Driven by a high concentration of motors, several new global features, or orders, emerge from a locally well-connected network. To quantify and even to forecast such new orders from an inhomogeneous system that is far from equilibrium, we converted the physical networks into mathematical graphs that reveal the pattern of a higher-order scaffold within the complex network. Using network sciences tools [21], we discovered a new type of avalanche in actomyosin networks related to a sudden change in the topology of a well-connected actomyosin network.

We then introduced these new features to train machine learning (ML) [22-25] models for forecasting these interesting far-from-equilibrium events. As an exploration of the predictability of avalanches, we trained two supervised machine learning models (support vector machine [26] and XGBoost [27]) with only the mechanical description of the actin filaments. The latter follows a gradient tree-boosting algorithm that is much more sophisticated than the former that follows a simple linear regression algorithm. In ML and supervised ML in particular, data curation and feature extraction are crucial for building reliable prediction models. We used features with physical interoperation from both polymer and network theory.

We utilized the two representative supervised machine learning models to explore order parameters for feature learning. We considered three polymer physics order parameters (the mean filament displacement, the radius of gyration, and the shape) [13] and six network theory order parameters (the density, the average clustering, the clique number, the mean closeness, the mean betweenness, and the assortativity) [21] for feature learning. Not only have we forecasted avalanches with great high probability, but we have also shown that the avalanches are mechanically dominated rather than chemically in the actomyosin network. The consideration of the features from the network theory order parameters into the training greatly improved the performance of both machine learning models by minimizing false negatives, benefiting the support vector machine model more than the XGBoost model. Our work has greatly expanded the toolset available for analyzing or interpreting protein-mediated actomyosin networks.

## II. Methods

### 1. Mechanochemical Dynamics of Active Networks (MEDYAN)

We simulated the dynamics of actomyosin networks by using a coarse-grained mechanochemical model of active systems called Mechanochemical Dynamics of Active Networks (MEDYAN), developed by the Papoian group [15, 20, 28-31]. The highlight of MEDYAN is its inclusion of mechanochemical feedback of the active networks, which makes the software more appropriate for our research interests compared to the models used in previous studies. MEDYAN consists of four main steps in its mechanochemical loop, as described in Figure 1.

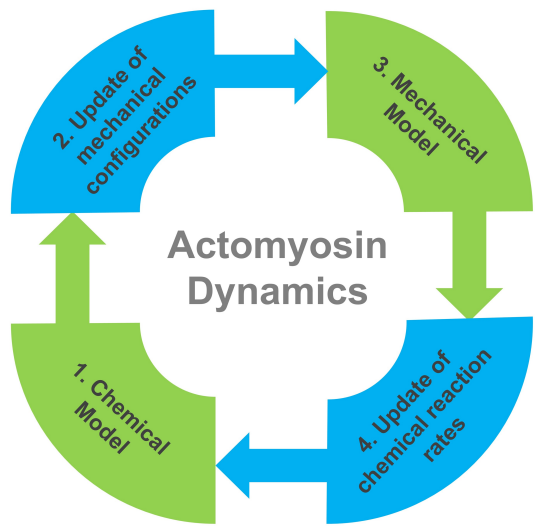

**Figure 1. The workflow of MEDYAN consists of four steps:** 1. Initiate the chemical model: The system evolves the actomyosin network with stochastic chemical reaction diffusion. 2. Updating the mechanochemical configurations: The chemical reactions deform the network locally, followed by the formation of a new mechanical configuration. 3. Initiate the mechanical model: The total energy of the new mechanical configuration is minimized with the conjugate gradient method by reaching a new equilibrium. 4. Update the chemical reaction rates: The chemical reaction rates are mechanochemically updated at the new equilibrium state. These four steps are cycled through for the entirety of the simulations. For a detailed description of MEDYAN, please see reference [15].

### 2. Simulation settings

#### 2.1 Simulation parameters

All simulations were confined to a three-dimensional, 1 μm³ rigid cubical box to match the size of a typical dendritic spine. The maximum time of the simulation is set to 2,000 seconds, and snapshots are captured every 10 seconds. Initially, the number of actin filaments was 50, and the filament length was 10 monomers. An example of typical snapshots of the simulations is shown in Figure 2, visualized with Mayavi 4.7.0 [32].

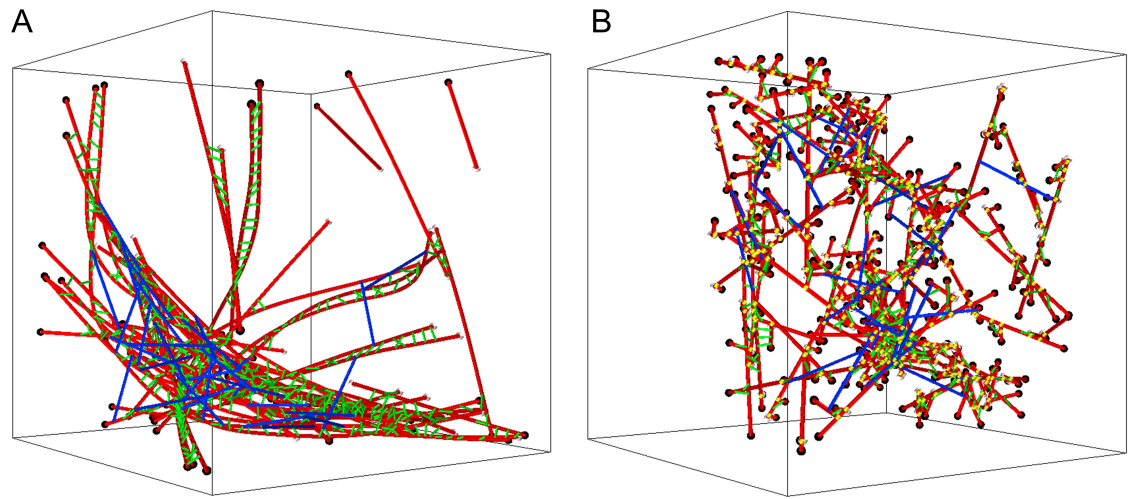

**Figure 2. Typical snapshots of MEDYAN simulations for the unbranched actomyosin networks without Arp2/3 complexes (A) and for the branched actomyosin networks with Arp2/3 complexes (B).** In both snapshots, a red cylinder represents an F-actin filament; a black bead represents a plus end of an F-actin filament; a white bead represents a minus end of an F-actin filament; a blue cylinder represents an ensemble of NMIIA motors that consists of 15 to 30 motor heads; a green cylinder represents an α-actinin cross-linker; and a yellow bead represents an Arp2/3 complex.

## 2.2 Reaction rates and mechanical constants

The reaction rates used in the simulations are listed in **Table 1**. $k_{p+}$ or $k_{p-}$: polymerization reactions of F-actin on the plus ends or minus ends; $k_{dp+}$ or $k_{dp-}$: depolymerization reactions of F-actin on the plus ends or the minus ends; $k_{bl}$ or $k_{ubl}$: binding or unbinding reactions of α-actinin linkers; $k_{bm}$ or $k_{ubm}$: binding or unbinding reactions of NMIIA motors; $k_{wm}$: the walking reactions of NMIIA motors; $k_{bf}$: the branching reaction of F-actin; $k_{df}$: the destruction reaction of a short F-actin no longer than one segment. The branching or destruction reactions are included only in branched simulations. For the details of chemical reactions in MEDYAN, please refer to [13, 15].

**Table 1: The reaction rates in the chemical model of MEDYAN**

| | Reaction rates | Value |
|---|---|---|
| Actin filaments | $k_{p+}$ | 0.151 s$^{-1}$ [15] |
| | $k_{p-}$ | 0.017 s$^{-1}$ [15] |
| | $k_{dp+}$ | 1.4 s$^{-1}$ [33] |
| | $k_{dp-}$ | 0.8 s$^{-1}$ [33] |
| Linkers | $k_{bl}$ | 0.009 s$^{-1}$ [34] |
| | $k_{ubl}$ | 0.3 s$^{-1}$ [34] |
| Motors | $k_{bm}$ | 0.2 s$^{-1}$ [35] |
| | $k_{ubm}$ | 1.7 s$^{-1}$ [15] |
| | $k_{wm}$ | 0.2 s$^{-1}$ [15] |
| Branching | $k_{bf}$ | 0.0001 s$^{-1}$ |
| Destruction | $k_{df}$ | 1.0 s$^{-1}$ (only applied to actin filament with 1 segment) |

The mechanical constants used in the simulations are listed in **Table 2**. $k_{\text{bend}}$: the filament bending constant; $k_{\text{stretch}}$: the filament stretching constant; $k_{\text{volume}}$: the volume force constant; $k_{\text{motor}}$: the motor stretching constant; $k_{\text{linker}}$: the cross-linker stretching constant; $k_{\text{boundary}}$: the boundary constant; $\lambda$: the boundary screening length constant; $k_{\text{stretch}}^{\text{branch}}$: the branching point stretching constant; $k_{\text{bend}}^{\text{branch}}$: the branching point bending constant; $\theta_0$: the branching point bending angle; $k_{\text{dihedral}}^{\text{branch}}$: the branching point dihedral constant; $k_{\text{position}}^{\text{branch}}$: the branching point position constant. For detailed force field definitions in MEDYAN, please refer to [13, 15].

**Table 2: The mechanical constants in the mechanical model of MEDYAN.**

| | Mechanical constants | Value |
|---|---|---|
| Actin filaments | $k_{\text{bend}}$ | 2690 pN·nm |
| | $k_{\text{stretch}}$ | 100 pN/nm |
| | $k_{\text{volume}}$ | 100000 pN·nm$^4$ |
| Motors | $k_{\text{motor}}$ | 2.5 pN/nm |
| Linkers | $k_{\text{linker}}$ | 8.0 pN/nm |
| Boundary repulsion | $k_{\text{boundary}}$ | 41 pN·nm (10 k$_B$T) |
| | $\lambda$ | 2.7 nm |
| Branched filament | $k_{\text{stretch}}^{\text{branch}}$ | 100 pN/nm |
| | $k_{\text{bend}}^{\text{branch}}$ | 100 pN·nm |
| | $\theta_0$ | ~70° [3, 4] |
| | $k_{\text{position}}^{\text{branch}}$ | 100 pN·nm |

**2.3. The setting of actin-binding protein concentration**

To explore the extent of actin-binding proteins on actomyosin dynamics, we chose several concentration ratios of actin-binding proteins to total actin. Five sets of motor and linker concentrations were selected to replicate the *in vitro* experiments from the Weitz group [18], while the three brancher concentrations were selected to investigate the impact of Arp2/3 concentration on the dynamics of the actomyosin network, as shown in Table 3. For referring to the Arp2/3 concentrations in the study, we refer to $x_{b:a}$ = 0.002 as the low brancher concentration, $x_{b:a}$ = 0.02 as the medium brancher concentration, and $x_{b:a}$ = 0.2 as the high brancher concentration.

**Table 3. The five sets of concentration ratios of motors or linkers to actin and the three concentration ratios of branchers to actin.** $x_{m:a}$ represents the ratio of motor concentration to actin concentration. $x_{l:a}$ represents the ratio of the $\alpha$-actinin cross linker concentration to the actin concentration. $x_{b:a}$ represents the ratio of brancher (Arp2/3) concentration to actin concentration.

| $x_{m:a}$ | $x_{l:a}$ | Motors | Linkers |
|---|---|---|---|
| 0.01 | 0.01 | Low | Low |
| 0.01 | 0.5 | Low | High |
| 0.5 | 0.01 | High | Low |
| 0.05 | 0.1 | Medium | Medium |
| 0.5 | 0.5 | High | High |

| $x_{b:a}$ | Branchers (i.e., Arp2/3) |
|---|---|
| 0.002 | Low |
| 0.02 | Medium |
| 0.2 | High |

# 3. Data analysis

## 3.1 Polymer physics order parameters.

Three polymer physics order parameters, the radius of gyration ($R_g$), the mean displacement of filaments ($\delta x_F$), and the shape parameter ($S$), are used to describe the macroscopic properties of the system. Their definitions are in SI, equations (S1-S4).

## 3.2 Network theory order parameters

Network theory is utilized in this study to capture the hidden properties of the actomyosin network. Our group has previously implemented network theory to characterize the complex topology in actomyosin dynamics. Our work reveals hidden properties involving uneven changes in the shape or the size of a network that are not captured by conventional order parameters [21].

We followed the steps to build a mathematical graph, $G(V, E)$, from physical actomyosin networks based on the proximity map of actin filaments. A proximity map is a matrix determined from the positions of actin filaments, where the coordinates of the plus end of each actin subunit are recorded as the position of a node ($V$). We chose 20 nm as the cutoff distance for constructing the proximity map, if the distance between a pair of nodes is less than 20 nm, it is assigned an edge ($E$) on the graph. In order to profile the topological arrangement that evolves into hierarchical, higher-order complexes in a network of actomyosins, we opted for tracking the filaments that are generally in close contact but not in direct contact by ignoring the chemical connectivity of filaments formed by actin-binding proteins such as a motor or a linker. The lengths of motors and linkers are 200 ± 25 nm and 35 ± 5 nm, respectively [36, 37]. We also do not include the two adjacent actin monomeric units in a filament measured at 27 nm. Therefore, the choice of short cutoff at 20 nm in constructing a proximity map will satisfy the purpose of tracking the emergent features of a complex network, while not capturing the chemical connectivity of actin.

In this way, we converted the actomyosin networks into mathematical graphs. The conformation of the mathematical graph from the physical actomyosin network and the calculation of network theory order parameters were performed with the Python package NetworkX [38]. Detailed descriptions and definitions of the parameters are shown below, as well as in SI equations S5-S11.

Graph ($V, E$): The components of a graph $G$ ($V, E$) include $V$: a set of vertices (also called nodes or points, filled circles in **Figure 3**) and $E$: a set of edges that connect the vertices (black solid lines in **Figure 3**). **Figure 3A** shows an example of a graph composed of 6 nodes and 8 edges. The degree of a node is the number of edges that are connected with it. For example, the degree of node *a* is 1, while the degree of node *b* is 4.

Assortativity: **Figure 3B** shows an example graph that has lower assortativity than the graph in **Figure 3A**. Assortativity, $\rho$, measures the tendency of nodes with similar degrees to be directly connected. The graph in **Figure 3A** has most of the nodes with similar degrees (b, c, d and e) connected to each other, while the nodes with similar degrees in the graph in **Figure 3B** are not connected. Therefore, the graph in **Figure 3A** has higher assortativity than the graph in **Figure 3B**. We employed the Z-score of the time derivative of $\rho$, $Z_{\Delta\rho}$ as one of the order parameters to probe undetected changes in the patterns. $Z = \frac{\rho - \mu}{\sigma}$, $\mu$ is the mean and $\sigma$ is the standard deviation of $\rho$.

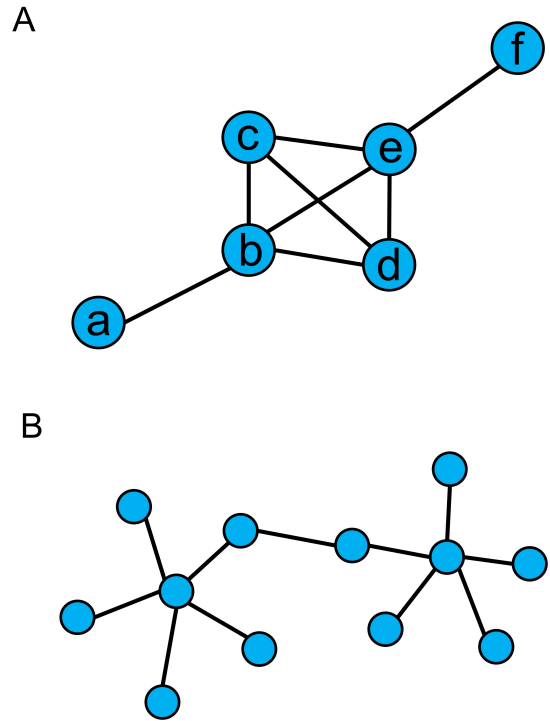

**Figure 3. Illustrating graph properties with examples.** (A) shows an example graph with high assortativity. (B) shows an example graph that has lower assortativity than the graph in (A).

### 3.3.1 Machine learning models to forecast avalanches

To predict avalanches in the mesoscopic simulations of actomyosin dynamics, we adopted two types of supervised learning models for training the dataset simulated with MEDYAN software: the SVM (support vector machine) model [26] and the XGBoost model [27]. Although both are widely used in machine learning, the XGBoost model has been proven to be more sufficient (better performance) and less expensive (consuming less computing resource) than linear regression models in most cases, while the performance of the SVM model depends highly on the choice of the kernels [39-41]. We trained the data with a polynomial kernel in the SVM model provided by the python scikit-learn 0.23.2 package [26]. For the XGBoost model, we trained the data using the python XGBoost 1.3.0 package [27].

The dataset used for training consists of 335 snapshots that precede avalanches as a positive dataset ("avalanche"), while for a negative dataset ("no-avalanche"), we used 493 other snapshots that were not succeeded by an avalanche. All these avalanches were selected from the MEDYAN simulations with the parameters of $x_{b:a} = 0.02$ and $x_{m:a} = 0.5$, representing the actomyosin dynamics at a medium brancher concentration and a high motor concentration, a condition favorable for the avalanches because of its rich connections nucleated by the branchers and abundant forces generated by the motors.

We have justified the upper threshold of 50 nm in the mean filament displacement ($\delta x_F$) for the selection of avalanche events from the snapshots of trajectories. We selected the individual snapshots with $\delta x_F$ over the upper threshold and assigned them to the positive dataset for training machine learning models. We visually inspected these snapshots that indeed there is a structurally large change over a short period of time. We also selected snapshots with a mean filament displacement less than the lower threshold of 20nm and assigned them into a negative dataset for machine learning models. We confirmed them (without an avalanche) by visual inspection. We have included more details in **Figure S5** in the supplementary information.

For each snapshot in both the positive and negative the dataset, we computed nine order parameters, or features, to characterize the morphologically complex structures in an actomyosin network. There are three from polymer physics: the radius of gyration ($R_g$), the mean displacement ($\delta x_F$), and the shape parameter ($S$). There are an additional six from network theory: the density, the clique number, the average clustering, the mean betweenness, the mean closeness, and the assortativity ($\rho$). The total dataset of 335 positives and 493 negatives is selected into two parts by the Python scikit-learn package [26]: 60% for training the model and 40% for testing the performance of the model. The training and testing sets, training_data.csv and testing_data.csv, are provided in SI.

### 3.3.2 Data analytics for the machine learning models

**Quality indicators**: A true positive (TP)/true negative (TN) is an outcome where the model correctly predicts the positive/negative class. A false positive (FP)/false negative (FN) is an outcome where the model incorrectly predicts the positive/negative class. The true positive rate (TPR), false positive rate (FPR), precision, or recall are defined below.

$$\text{TPR} = \text{TP}/(\text{TP} + \text{FN}) \hspace{4cm} (\text{Eqn. 1})$$

$$\text{FPR} = \text{FP}/(\text{FP} + \text{TN}) \qquad \text{(Eqn. 2)}$$

$$\text{Precision} = \text{TP}/(\text{TP} + \text{FP}) \qquad \text{(Eqn. 3)}$$

$$\text{Recall} = \text{TP}/(\text{TP} + \text{FN}) \qquad \text{(Eqn. 4)}$$

With such quality indicators, we measured the performance of the machine learning models by using the receiver operating characteristic curve, the precision-recall curve, the area under the curve, and the confusion matrix as defined below:

Receiver operating characteristic (ROC) curve: The ROC curve was created by plotting the true positive rate (TPR) against the false positive rate (FPR) at various threshold settings randomly selected by the Python scikit-learn package.

Precision-recall (PR) curve: The PR curve is created by plotting the precision against the recall at various threshold settings randomly selected by the Python scikit-learn package.

The area under the curve (AUC): the ratio of the area under the curve to the total area in a figure. The AUC ranges from 0 to 1. A larger AUC value indicates a better performance of a model.

Confusion matrix: A confusion matrix is used to present the details in the performance of the models. The confusion matrix is composed of true class in rows and predicted class in columns. The numbers in the table are the numbers of the four quality indicators (TP, TN, FP, and FN) described above. A confusion matrix with higher true positives (TPs) and true negatives (TNs) indicates a more accurate machine learning model.

### III. Results

**1. The content of the Arp2/3 complex modulates the dynamics of actomyosin networks**

To explore the influence of the Arp2/3 concentration that impacts the content of branched networks in the contractility of an actomyosin network, we compared the $R_g/R_g^i$ time courses of actomyosin systems with low, medium and high brancher concentrations, as shown in **Figure 4 A, B, and C,** respectively. $R_g$ is the radius of gyration, and $R_g^i$ is the initial condition of $R_g$. The increase and decrease in $R_g/R_g^i$ over time indicate the expansion and contraction of the actomyosin network.

At low Arp2/3 concentrations ($x_{b:a}$ = 0.002) in **Figure 4A**, $R_g/R_g^i$ varies broadly with motor and linker concentrations. At low motor and linker concentrations ($x_{m:a}$ = 0.01, $x_{l:a}$ = 0.01, thin solid line in **Figure 4A**), $R_g/R_g^i$ initially increases over time, reflecting the expansion of the networks. When motor or linker concentrations both increased ($x_{m:a} > 0.01$ and $x_{l:a} > 0.01$), the contractility of actomyosin networks reacted differently by either the active (i.e., motors) or passive actin-binding proteins (i.e., linkers). $R_g/R_g^i$ shows that an increase in motor concentration always promotes contraction of an actomyosin network (dotted line compared to thin solid line, thick solid line compared to dashed lines in **Figure 4A**, $x_{m:a}$ = 0.5 compared to 0.01). Meanwhile, either a high or low linker concentration leads to a similar $R_g/R_g^i$ over time at a steady state (thick solid line and dotted line in **Figure 4A**, $x_{l:a}$ = 0.5 and 0.01). These observations are consistent with the findings from prior experimental[18] and theoretical investigations[13].

$R_g/R_g^i$ varies less over time when we increase the ratio of the Arp2/3 concentration over actin to 0.02 (**Figure 4B**), but the profiles are still modulated by the motor and linker concentrations. When the ratio is increased to the highest level at 0.2 (**Figure 4C**), $R_g/R_g^i$ does not significantly change over time regardless of the motor and linker concentrations. The dynamics in these networks slow down significantly. The actomyosin network remains highly branched and static after the motors are unable to walk along the filaments saturated with Arp2/3 complexes.

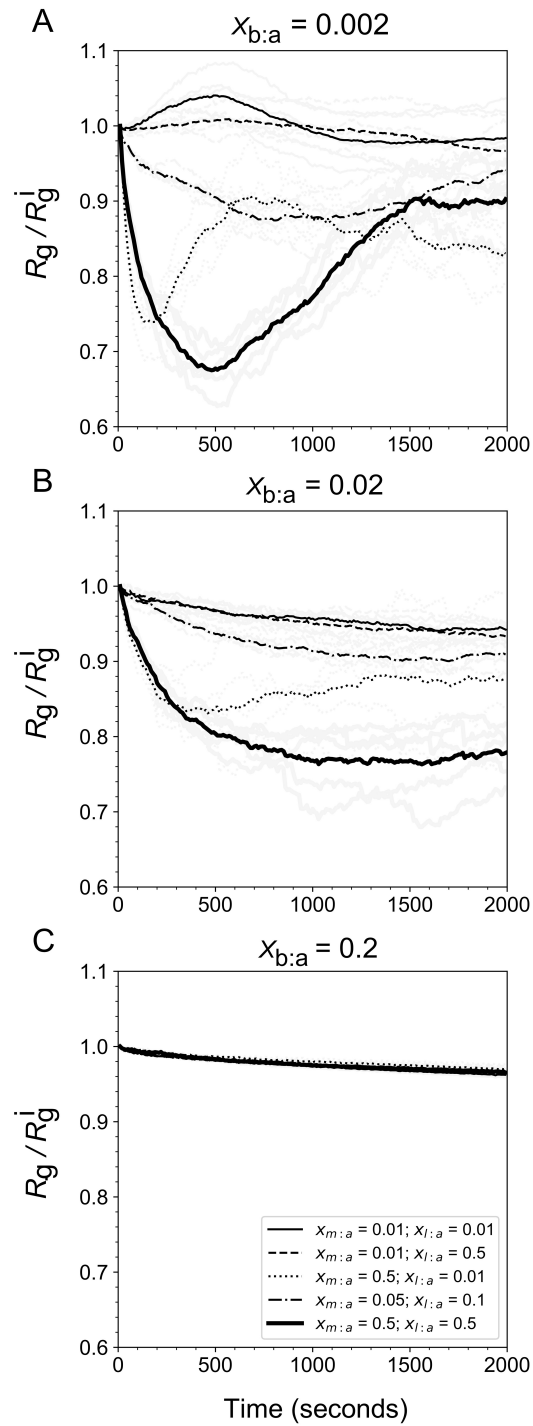

**Figure 4. Time courses of $R_g/R_g^i$ in branched networks with low, medium, and high brancher concentrations.** The black lines in A, B and C show the time courses of $R_g/R_g^i$ of simulations with different motor and linker concentration pairs under the conditions of low ($x_{b:a}$=0.002), medium ($x_{b:a}$=0.02) and high brancher concentrations ($x_{b:a}$=0.2), each averaged from five simulations with the same initial conditions (averaged results are shown in black, and the five original trajectory replicates are shown in light gray). Thin solid lines represent networks with low motor and linker concentrations ($x_{m:a}$ = 0.01, $x_{l:a}$ = 0.01); dashed lines represent networks with low motor and high linker concentrations ($x_{m:a}$ = 0.01, $x_{l:a}$ = 0.5); dotted lines represent networks with high motor and low linker concentrations ($x_{m:a}$ = 0.5, $x_{l:a}$ = 0.01); dashed dotted lines represent networks with medium motor and linker concentrations ($x_{m:a}$ = 0.05, $x_{l:a}$ = 0.1); and thick solid lines represent networks with high motor and linker concentrations ($x_{m:a}$ = 0.5, $x_{l:a}$ = 0.5). The definitions of $x_{m:a}$, $x_{l:a}$ and $x_{b:a}$ are described in Table 3.

## 2. Network theory facilitates data visualization by converting a physically complex network to a mathematical graph

We visualized snapshots of actomyosin networks at low, medium or high Arp2/3 concentrations while the motor and linker activities remained the same in **Figure 5A, C, and E**, respectively. Their structural morphologies are distinctive, and the tensions are distributed unevenly throughout a complex network. The length of the filaments at a higher Arp2/3 concentration (**Figure 5E**) are shorter, and the tension within them is lower compared to those at a lower Arp2/3 concentration (**Figure 5A**). When the Arp2/3 concentration was medium (**Figure 5C**), the filament lengths and tension degree were also at a medium level. We speculated that there is a causal relationship between the complexity of the physical network and the distribution of tensions, which dictates the emergent dynamics of actomyosin networks.

However, the first step is to quantify the complexity emerging from an actomyosin network by converting a physical network into a mathematically representative graph. This transformation reveals the hidden features of actomyosin networks in the MEDYAN simulations by filtering out unimportant information as noise. We visualized typical snapshots from the simulations with low ($x_{b:a}$=0.002), medium ($x_{b:a}$=0.02) or high ($x_{b:a}$=0.2) Arp2/3 concentrations in **Figure 5A, C, or E**, respectively, in mathematical graphs in Figure **5B, D, or F.** Indeed, these mathematical graphs are distinctive in both sizes and structures and reveal hidden features through the sizes and connectedness of a node (which represents an actin filament). At a local level, the number of nodes with high degrees increases with the concentration of Arp2/3. In addition, the mathematical graphs show that there are more connections between nodes (i.e., filaments) at higher Arp2/3 concentrations than those at lower

Arp2/3 concentrations.

The network theory order parameters excel at revealing hidden features at a nonlocal level, which is a challenging task in a nonhomogeneous system. We revealed the importance of these hidden features by measuring the "betweenness centrality" of a node (described in **SI 1.3.4**). When the Arp2/3 concentration is low, there are only a few nodes with high betweenness (shown by a large node) on the graph in **Figure 5B**, corresponding to a highly connected hub of actomyosin networks in **Figure 5A**. When the Arp2/3 concentration is medium (**Figure 5D**), several high-betweenness components emerge, corresponding to several delocalized and sparsely connected clusters within the physical network (**Figure 5C**). When the Arp2/3 concentration is high in **Figure 5F**, compared to the networks with low Arp2/3 concentration, there are more nodes with low betweenness emerge (shown by decentralized, small nodes). We interpreted that these nodes are spatially far apart from one another. They are categorized into several communities with different colors. The layout approach created by Yifan Hu is useful to visualize the community in a complex network by arranging the nodes hierarchically on a mathematical graph in **Figure 5B** and **5D**. Overall, there was no clearly defined central hub in the network (**Figure 5E**). Physically, these nodes are the individual mother filaments with their daughter filaments (**Figure 5E**), while the former are not connected to one another.

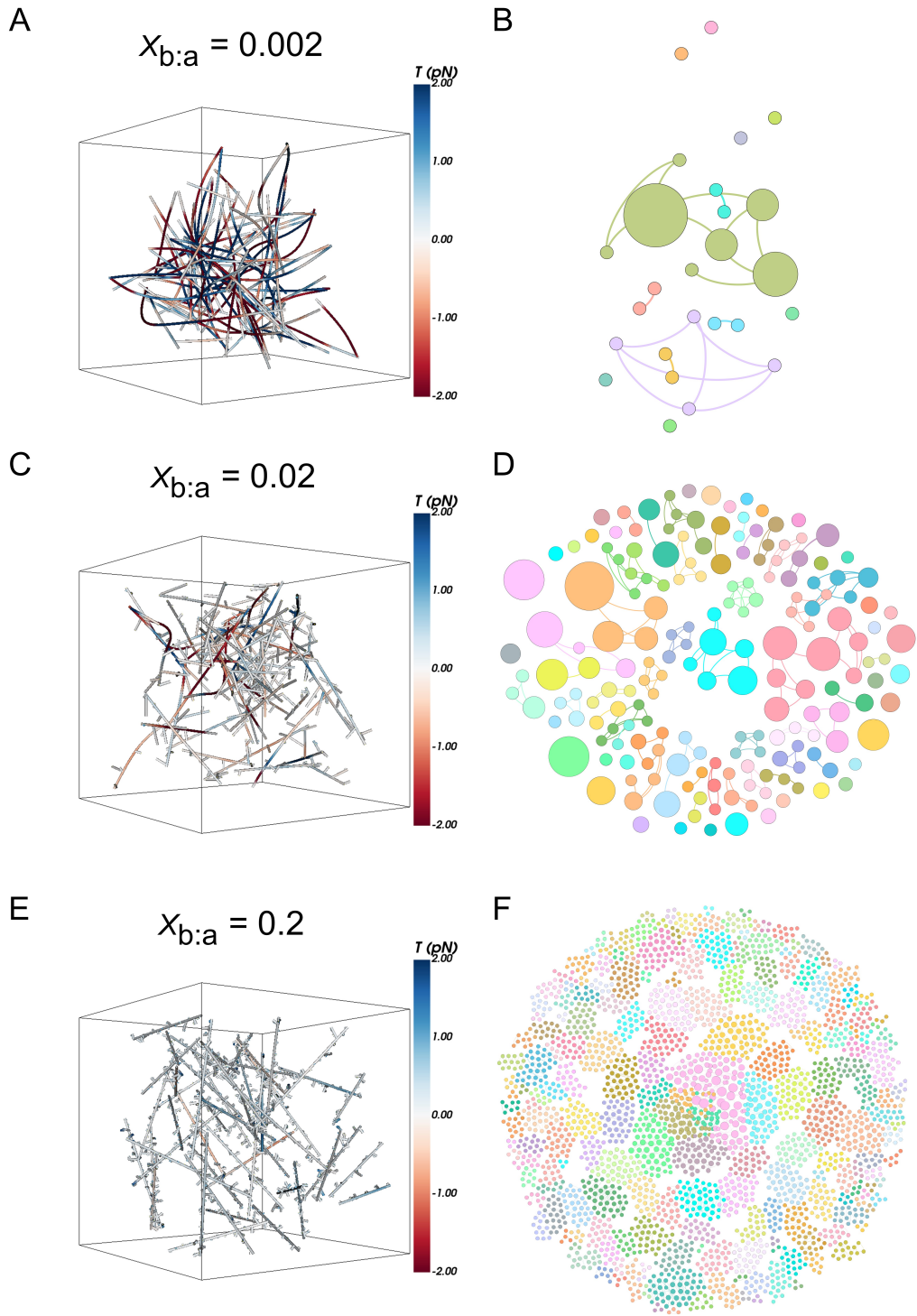

**Figure 5. Graph networks of physical actomyosin networks at several Arp2/3 concentrations at 500 s.** A, C, and E show snapshots of the simulations at low, medium and high arp2/3 concentrations, while they all have the same high motor and linker concentrations. The actomyosin filaments are colored by tension in $pN$, while motors, linkers, and branchers are not shown on the snapshot. A positive value of tension represents stretched filaments, while a negative value of tension represents compressed filaments. B, D, and F are graph representations of the actomyosin network from A, C, and E, respectively. We filtered out

nodes with a node degree less than 3 and the nodes with a self-loop for clear visualization. The layout algorithm by Yifan Hu with default parameters in Gephi 0.92[42] is used. The size of a node depends on its betweenness centrality, while the nodes are colored according to the identification number of a component (a component is a group of nodes that are connected inside the group but not connected to any nodes outside the group).

**3. Network theory order parameters reveal aster-like features from physically complex actomyosin networks at varying Arp2/3 concentrations**

At the community level, network theory order parameters such as assortativity, $\rho$, provide the outlook of the hierarchical structure within a network. The decrease in assortativity indicates the change of a network morphology from a 'centered' to an 'aster' topology, as shown in **Figure 3**. While we varied the Arp2/3 complex concentration in the simulations of actomyosin dynamics in **Figure 6**, we showed that an increase in Arp2/3 complex concentration leads to a decrease in assortativity, revealing an altered topology within a network to become aster-like.

Similarly, in **Figure 5E**, where the Arp2/3 complex concentration is high, we observed numerous short branches on the actin filaments, which is equivalent to the composition of having many 'aster-like' branches in the network (**Figure 5F**). The morphology of the actomyosin networks with high Arp2/3 concentrations is totally different from the actomyosin networks with low Arp2/3 concentrations, resulting from the formation of more short branches on the actin filaments mediated by Arp2/3 nucleation behavior.

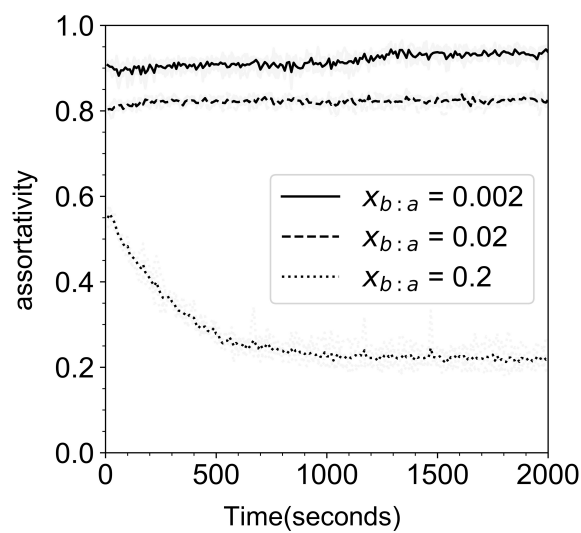

**Figure 6. Assortativity of networks with different Arp2/3 concentrations.** The figure shows the assortativity for actomyosin networks with low, medium and high brancher concentrations but the same high motor and linker concentrations. Solid lines represent networks with low Arp2/3 concentrations ($x_{b:a}$=0.002); dashed lines represent networks with medium Arp2/3 concentrations ($x_{b:a}$=0.02); dotted lines represent networks with high Arp2/3 concentrations ($x_{b:a}$=0.2). Each black line is the averaged result of five simulations with the same initial conditions (the five replicates are shown in light gray).

**4. The change in assortativity captures a new type of avalanche resulting from disruption in the hierarchical organization of an actomyosin network.**

An avalanche, or a cytoquake, is a sudden structural rearrangement of the networks captured by the positional changes of the filaments, $\delta x_F$, in simulations or experiments [13, 14]. In the computational investigation from our prior work [13], we have characterized two types of avalanche by employing the polymer physics order parameters such as the radius of gyration ($R_g$) and the shape parameter (S). $R_g$ and S measure distinctive properties of a network in terms of its size and shape, respectively. However, they do not capture the rearrangement of topology within a network that may neither distinctively impact the overall shape nor the size of a network. In this work, by varying the concentration of Arp2/3, we discovered another classification of avalanche caused by the hierarchical reorganization of a network (**Figure 7**). Such changes are structurally complex and mostly hidden by layers of information. We captured them with the aid of unique network theory order parameters.

We ranked the nine order parameters (three from the polymer physics and six from the network theory order parameters) by comparing their Pearson correlation against one another in **Figure S2**. Interestingly, $\rho$ is strongly anti-correlated with the other order parameters, signifying its importance in capturing emergent properties. Indeed, the change in $\rho$, $\Delta\rho$, is useful to probe the emergence of a higher-order organization from uneven distribution of local nodes.

We demonstrated the new characterization of the avalanches by showing a segment of the trajectory as an example in **Figure 7** that it is entirely different from the previous two kinds. The plot of $\delta x_F$ over time in **Figure 7** indicates three avalanches at 600, 680 and 960 seconds. We also showed the time derivatives of $R_g/R_g^i$, the shape parameter S, and the Z-score of time derivative of assortativity in **Figure 7A**, **7C** and **7E**, and provided the illustrations corresponding to the movements in **Figure 7B**, **7D** and **7F.** The avalanche at 600 seconds coincides with a sharp decrease in $\Delta(R_g/R_g^i)$ in **Figure 7A**, indicating that this is a contraction (**Figure 7B**). The avalanche at 960 seconds coincides with a large drop of $\Delta S$ (**Figure 7C**), signifying that the network deforms under shear and the shape becomes oblate (**Figure 7D**).

The avalanche at 680 seconds not only coincides with contraction and shape changes, but also uniquely coincides with a peak in $Z_{\Delta\rho}$, the Z-score of $\Delta\rho$ in **Figure 7E**, while the events at 600 seconds and 960 seconds do not. The sudden increase in assortativity at 680 seconds indicates that the topology of the network changes by altering the hierarchy of connected nodes. We illustrated this movement in **Figure 7F** and focused on the connectivity from the surrounding green nodes to node (a) and then to nearby node (b) during an avalanche. Once other nodes that connect to node (a) also connect to node (b), the node degree and the betweenness of node (b) grows. Consequently, the assortativity of the network increases by 0.11 because node (a) and node (b) that now share a similarly large node degree and high betweenness connect to each other (e.g., a tongue-in-cheek remark would be "guilt by association"). The connectivity of nodes changes from a 'aster-like' to a 'centered' topology. Although this new type of avalanche at 680 seconds may still carry changes in size and shape, it is an entirely new type of avalanche distinctive to the other two avalanches at 600 and 960 seconds.

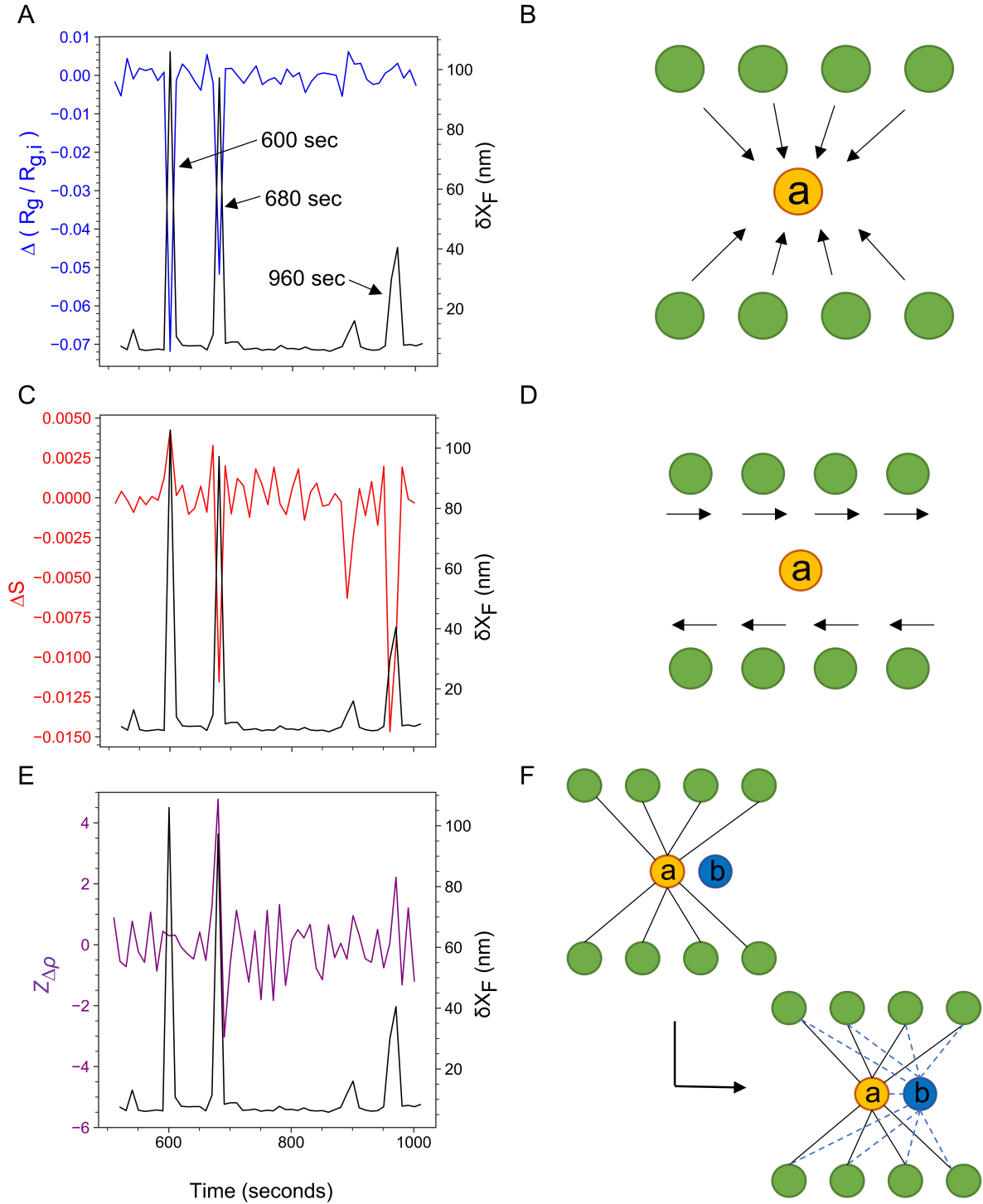

**Figure 7. Assortativity captures the third classification of avalanche.** $\delta x_F$ represents the mean displacement of filaments in the network. (A) $\Delta(R_g/R_g^i)$ represents the time derivative of the ratio of the current $R_g$ and the "initial" $R_g$ at 10 seconds, (C) $\Delta S$ represents the time derivative of the shape parameter, (E) $Z_{\Delta\rho}$ represents the Z-score of the time derivatives of assortativity, $\rho$, an order parameter measuring the topology of a network. (B), (D) and (F) are the cartoons that illustrate the changes in the size, shape and topology of a network,

corresponding to (A), (C) and (E), respectively. The black arrows in (B) and (D) represent the moving directions of the green nodes relative to node (a). The black lines in (F) represent the edges from the green nodes to node (a) before the avalanche. The blue dashed lines represent newly formed edges connecting other nodes to a nearby node (b) during the avalanche. An increase in $Z_{\Delta\rho}$ indicates an increase in assortativity of the network. Illustratively, node (a) and node (b) now form a 'center-like' cluster during the avalanche. The example simulation has high motor, low linker and medium brancher concentrations.

Once we discovered the third type of avalanche at 680 seconds (**Figure 7E**), we visualized the physical networks (**Figure 8A** and **Figure 8B**) on a mathematical graph in **Figure 8C** and **8D** before and after the avalanche, respectively. Then, we computed nonlocal order parameters, particularly the betweenness centrality, to detect hidden patterns in a complex network. After the avalanche at 680 seconds, we observed the emergence of clustered green nodes with high betweenness in **Figure 8D**, which did not exist before the avalanche (**Figure 8C**). The formation of this center with high-betweenness nodes increases the assortativity of the network, reflected as a peak in **Figure 7E**.

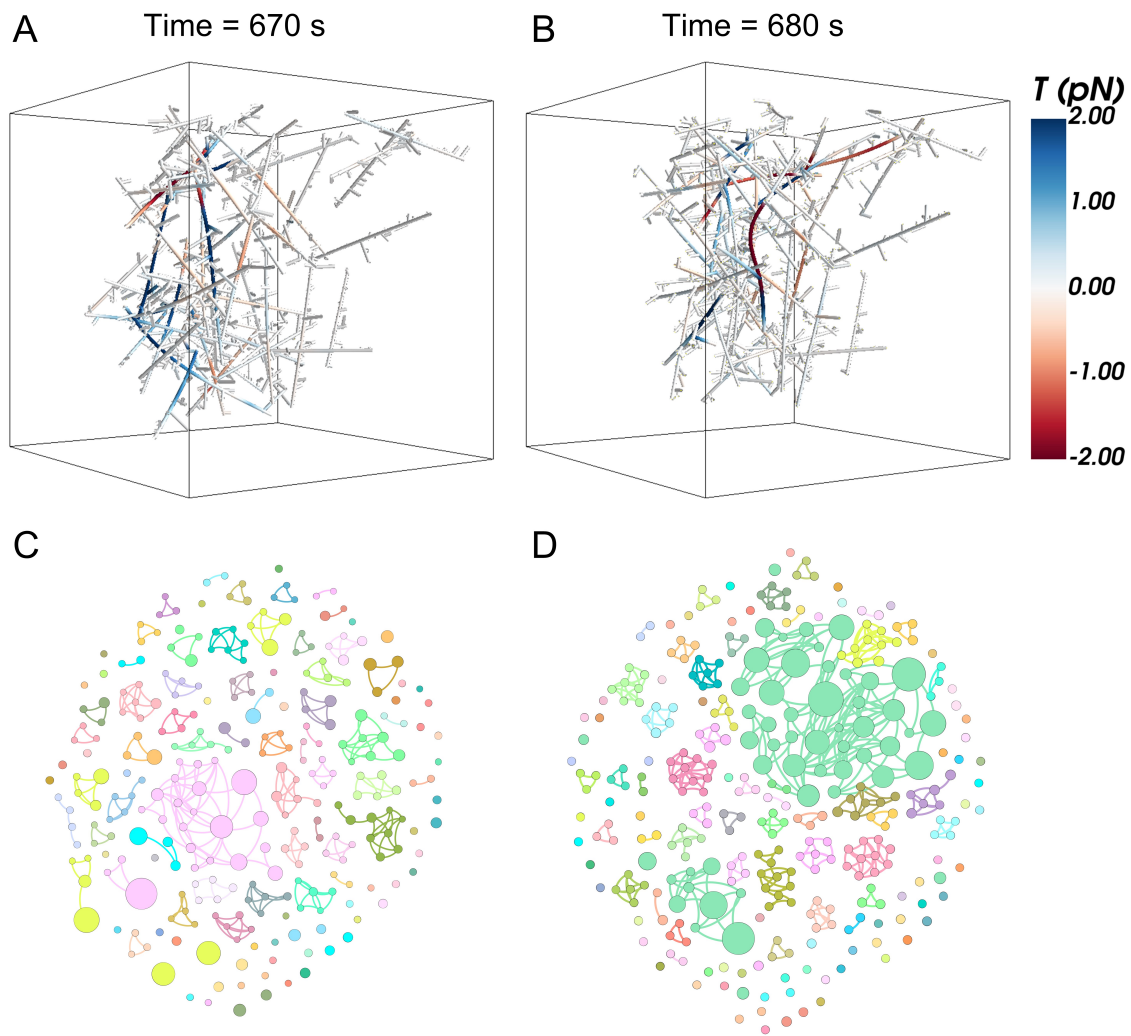

**Figure 8. Tension snapshots and corresponding visualized graphs for avalanches at 680 seconds.** A and B are the tension snapshots before (670 seconds) and during (680 seconds) the avalanche, and C and D are the corresponding visualized graphs of the snapshots. These two visualized graphs filtered out nodes with degrees lower than 6 and nodes with self-loops for clear output, and the layout algorithm Yifan Hu with default parameters in Gephi 0.92[42] was used during the visualization process. The size of a node depends on its betweenness centrality, while the nodes are colored according to component ID.

## 5. Machine learning tools were applied to forecast avalanches in actomyosin dynamics.

We compare the performance of machine learning (ML) models using the receiver operating characteristic (ROC) and precision-recall (PR) curves in **Figure 9**. Both ROC and PR curves provide a diagnostic tool for binary classification models for measuring the ability of a machine learning model to make correct predictions. The area under the curve (AUC) of the two curves provides quantitative scores that summarize the curves and can be used to compare classifiers. An AUC closer to 1 indicates a more skillful model [43-45].

Since we achieved a comprehensive understanding of the structural characteristics of avalanches, we will employ the order parameters as features to apply machine learning tools to forecast avalanches. While we compared two prominent machine learning (ML) models in predicting emergent avalanches in actomyosin networks, we also explored the importance of the features employed for training datasets.

For each machine learning model (XGBoost or SVM), we employed two sets of order parameters for training the dataset. The first set includes all nine order parameters: three polymer physics order parameters and six network theory order parameters. The second set includes only three polymer physics order parameters. As shown in **Figure 9A** and **9B**, all 9 parameters were used in training the model. For the SVM model (blue), the AUCs for the ROC curve and PR curve are 0.88 and 0.83, respectively. For the XGBoost model (red), the AUCs for the ROC curve and PR curve are 0.96 and 0.96, respectively. The XGBoost model performs better than the SVM model, which is expected since the XGBoost model is shown to be sufficient in most applied cases, while the performance of the SVM model relies strongly on the selection of a good kernel [39, 40]. Notably, the AUC values for the XGBoost model from our investigation are high enough (close to 1, the perfect value) to prove the excellent performance of this model in our case.

Next, we diagnosed the outcome with only 3 polymer physics order parameters used in training the models (**Figure 9C** and **9D**). For the SVM model (blue), the AUC for the ROC curve and that for the PR curve decrease significantly to 0.76 and 0.70, respectively, in comparison with those trained with 9 order parameters. For the XGBoost model (red), the AUCs for the ROC curve and the AUCs for the PR curve were 0.94 and 0.93, respectively. They remain at the same level as those trained with 9 order parameters, indicating that adding more features from network theory order parameters improves the performance of the SVM model more than the performance of the XGBoost model.

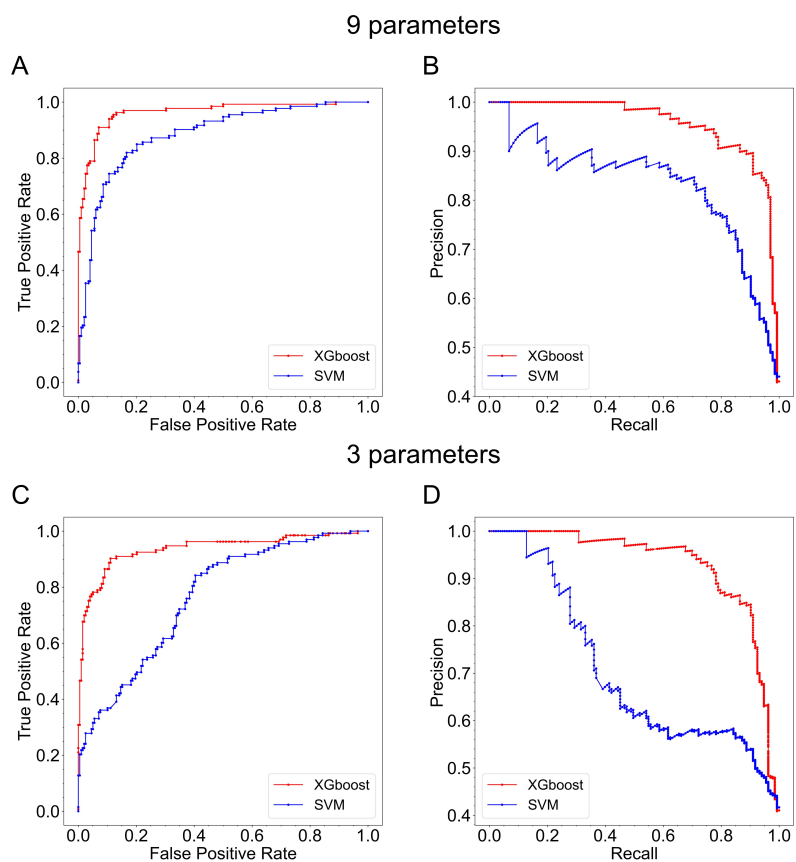

**Figure 9. ROC and PR curves for XGBoost and SVM models.** A and B show the performance of machine learning models trained with all 9 parameters: 3 polymer physics order parameters and 6 network theory order parameters. C and D show the performance of the models trained with only 3 polymer physics order parameters. A shows ROC curves for XGBoost and SVM models trained with all 9 parameters, B shows the PR curves for these two models; C and D show the ROC and PR curves of the same types of models trained with only 3 parameters.

## 6. Network theory order parameters strengthen machine learning models to forecast avalanches better in actomyosin dynamics

We further diagnosed the performance of the two ML models by employing the confusion matrices that label the true and predicted cases in **Figure 10**. The definition of a confusion matrix is explained in Methods 3.3.2. For the SVM model, the addition of the six network theory order parameters into the ML training dataset (**Figure 10A** and **10C**) moved 38 counts from the category of false negatives to the category of true positives, and 1 count of false positives to one count of true negatives. Meanwhile, for the XGBoost model, the addition of the six network theory order parameters into the ML training dataset (**Figure 10B** and **10D**) brings only 8 counts from the category of false negatives to that of true positives and 3 counts from the category of false positives to the category of true negatives. The addition of network theory order parameters into the ML model datasets enhances the forecast of avalanches, especially by reducing the number of false negative predictions, indicating more predicted hidden avalanches.

The XGBoost model exhibits better performance than the SVM model with both three and nine parameter datasets, indicating that the XGBoost model is potentially a better classifier in our case for the prediction of avalanches. Additionally, the XGBoost model shows less sensitivity to the network theory order parameters than the SVM model. These facts motivate us to further investigate how these nine features work during the training of this model. Therefore, we further evaluated the importance of these features in the XGBoost models in **Figure 11**.

As shown in **Figure 11A**, when the XGBoost model was trained with 9 features, the three polymer physics order parameters mean displacement, shape and radius of gyration have a prior importance in avalanche prediction. Most of the network theory order parameters have secondary importance with the exception of the mean betweenness. As discussed in the mathematical graphs of **Figures 5 and 8**, the betweenness of nodes tracks the centrality distributions in the network, measuring the 'shortest pathways' on actin filaments in a physical network. As a mechanical emergent phenomenon in actomyosin networks, avalanches are closely related to the formation and subsequent delocalization of clustered centers in the physical network. Therefore, the mean betweenness plays a more important role than other network theory order parameters in forecasting avalanches.

**Figure 11B** shows that in the XGBoost model trained with only three polymer physics order parameters, the importance of these three parameters is relatively similar, which is consistent with **Figure 11A**. In contrast, when we used only the 6 network theory order parameters to train the dataset, the forecast of an avalanche was poor (see **SI Figure S4**). The ensemble of decision trees[27] in XGBoost is better at detecting hidden patterns from a complex network than linear regression in the SVM model. Although introducing network order parameters in training the dataset improves the performance of both models, the extent of performance is more significant for the SVM model than for the XGBoost model.

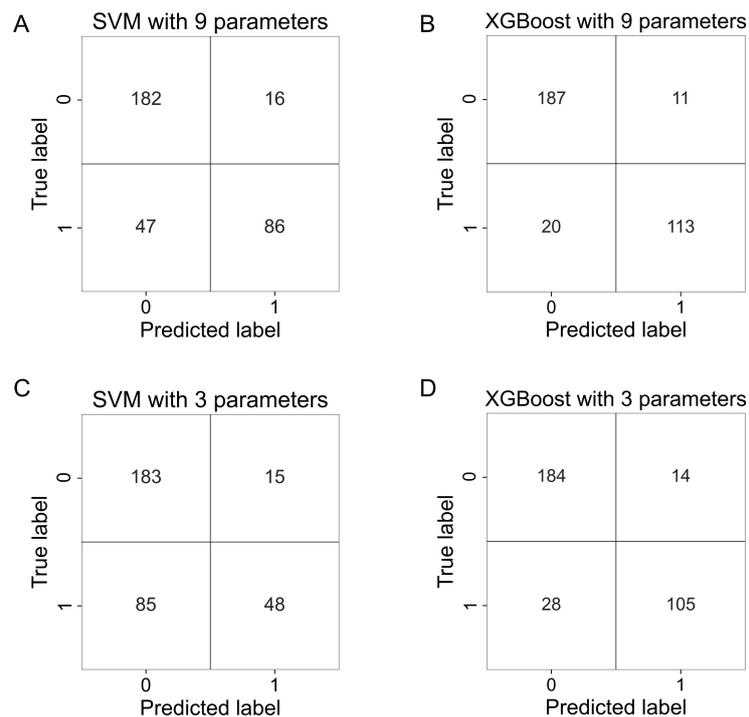

**Figure 10. Confusion matrices for XGBoost and SVM models.** A and B show the confusion matrices of the XGBoost and SVM models trained with all 9 parameters: 3 polymer physics order parameters and 6 network theory order parameters. C and D show the confusion matrices of the XGBoost and SVM models trained with only 3 polymer physics order parameters. In the confusion matrices, label 1 represents a snapshot that is followed by avalanche while label 0

represents a snapshot that is not followed by avalanche.

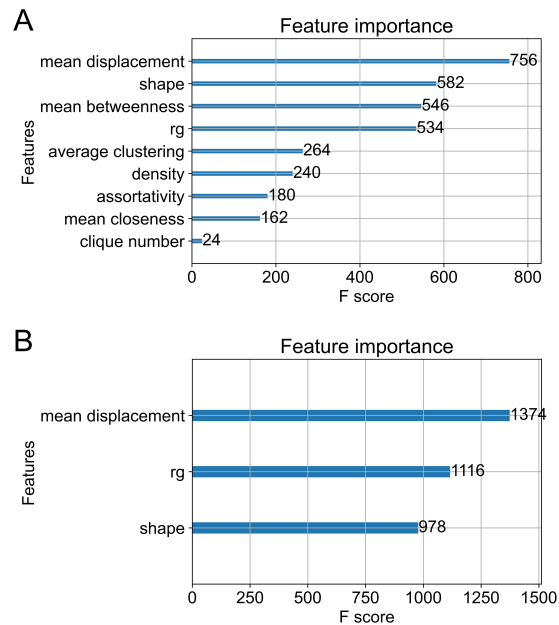

**Figure 11. The feature importance of parameters in the XGBoost models.** A and B show the feature importance of the parameters used in the two XGBoost models trained with 9 and 3 parameters. A and B count the number of times each feature is split across all boosting rounds (trees) in the model as the F-score and visualize the result as a bar graph, with the features ordered according to how many times they appear.

**Discussion**

**I. The Arp2/3 complex concentration tunes the emergence of avalanches in actomyosin networks**

The actin-related proteins 2/3 (Arp2/3) complex, also known as the brancher in our system, initiates a filament branch (daughter filaments) at an angle of 70° on the sides of the preexisting mother, subsequently altering the topology of the network [4, 5]. The binding of the Arp2/3 complex to a filament is an ATP dependent process [46] to prevent passive unbinding. This piece of experimental evidence shows that it is exceedingly rare for the Arp2/3 complexes to unbind itself from actin filaments without an ATP-consuming reaction involving another enzyme. It motivates our reasoning of the parameterization of the Arp2/3 complex in which the event of Arp2/3 unbinding from actin filaments is quite rare in the MEYDAN simulations[13]. In our prior investigation, we justified the use of MEDYAN over other codes, such as Cytosim[7] or AFiNES[8] because MEDYAN[15] has physically realistic models and mechanochemical feedback, which is critical to describe active processes.

Here, our study shows that a high concentration of the Arp2/3 complex limits linker binding and motor walking, which in turn reduces connectivity and inhibits contraction of the network (**Figure 4C**). However, by decreasing the Arp2/3 concentration in the system, the network not only contracts faster and more robustly (**Figure 4A**) but also has the ability to form larger clusters at the center (**Figure 5A**). At an intermediate concentration of the Arp2/3 complex and a high concentration of motor proteins, the structure of actomyosin is marginally stabilized; thus, the "avalanche" phenomenon is most likely to occur (**Figure 4B and 5C**).

As the main part of the cytoskeleton, actomyosin networks play important roles in various cell behaviors. As an essential actin-binding protein, the distribution of Arp2/3 complexes has been experimentally proven to be related to cell motility in non-muscle cells [47, 48]. In addition, the branched actin network is especially rich at the edge of cells, such as the actin cortex, indicating the importance of this protein to the modulation of cell shape changes[2, 49]. This modulation is achieved by the treadmilling of branched networks nucleated by the Arp2/3 complex and other actin-binding proteins that sever filaments such as cofilin[50-52]. The impact of the Arp2/3 complex concentration on the simulated network structure and dynamics described in this study will advance our understanding of the role of the Arp2/3 complex in these cell behaviors.

## II. Network theory reveals a new type of avalanche associated with topological changes in a physical network

With the concentration of branchers (Arp2/3 complexes) dictating the nanostructure of the actomyosin network, which in turn alters the entire network topology, a proper tool is needed to describe the hierarchical properties of the system. Therefore, we chose network theory to analyze the simulated networks in this study. Mathematical graphs and the tools of data science prove to be superior in detecting hidden patterns within a complex network[21]. Order parameters such as clustering coefficient, betweenness and closeness measure the microscopic network properties down to a single actin subunit and characterize its role in nonlocal features, while order parameters such as assortativity and density measure the macroscopic properties of the whole network. Therefore, the network theory order parameters provide the needed mesoscopic descriptors that connect the microscopic properties to macroscopic phenomena in an active system far from equilibrium.

In particular, we discovered that betweenness is most useful for visualizing the connection between the microscopic and the macroscopic network properties when the Arp2/3 concentrations vary (note: Arp2/3 initiates branching). For example, the graphs in **Figure 5B, 5D and 5F** have distinguished betweenness distributions. In the mathematical graph of **Figure 5B**, only one community of nodes with the highest betweenness in large circles are connected to each other, while other communities of nodes with a much lower betweenness in small circles are gathered. It underpins a centralized cluster of actomyosin filaments at low Arp2/3 concentration in the snapshot shown in **Figure 5A**. In the mathematical graph of **Figure 5D**, several communities have higher betweenness (in larger circles) than other communities (in smaller circles). It indicates the presence of multiple centers in the actomyosin network in **Figure 5C**. In the mathematical graph of **Figure 5F**, the betweenness of nodes is small among most communities. It indicates delocalized communities in the actomyosin network (**Figure 5E**). The causal relationship between the size of communities and the value of betweenness supports the hypothesis that a higher brancher (i.e., Arp2/3) concentration leads to a global network with delocalized communities, involving significant changes in the rearrangement of network topology.

This new type of avalanche is related to the change of the network topology in a branched actomyosin network. We discovered these subtle changes in the network topology by observing the betweenness from a mathematical graph at the onset of an avalanche at 680 seconds (**Figure 8**). There is an increase in the number of nodes with high betweenness during the avalanche (**Figure 8D**) compared to that before the avalanche (**Figure 8C**). The changes in the betweenness indicate the altered connectivity from an aster-like to a centered-like node (**Figure 7E**). The sudden creation of distinctive communities in a network initiates an avalanche. We further revealed the hidden hierarchy of the network with assortativity and captured a new type of 'avalanche' involving the reorganization of a network from a 'delocalized' community to a 'centralized" one (**Figure 8**). In cells, actomyosin networks may have similar size and shape but distinguished intra-network topology. The emergence of new higher order risen above layers of actomysin filaments probably leads to distinct functions. Therefore, it is important to utilize assortativity or other order parameters in the network theory to reveal the hidden topological features among these networks.

**III. Forecast avalanches in actomyosin dynamics with network science and machine learning**

In ML and supervised ML in particular, data curation and feature extraction are crucial steps for building reliable prediction models. For forecasting avalanches, **Figure 9** shows how adding network theory parameters to SVM models increases their specificity and sensitivity, whereas XGBoost models do not have this strong impact. As a rule of thumb, XGBoost is a suitable option, especially for small datasets such as ours as compared to other machines learning models [39-41]. However, the SVM model is a naiver approach since it is merely a linear regression model that relies heavily on selecting the right kernels and lacks the ability to boost the model multiple times with the same dataset.

A key feature is the betweenness centrality that captures the formation and disappearance of a cluster from a complex network. We showed that adding the betweenness centrality into the training set with the polymer physics order parameters greatly increases the performance of SVM, while other network theory order parameters are of secondary importance in **Figure 11.**

We believe that our approach can also be used in predicting an avalanche in experiments where the positions of actin filaments are easier to track than actin-binding proteins. The discovery of hidden patterns can be achieved by converting a physical network into a mathematical graph and the forecasting of avalanches can be predicted by ML.

### IV. Conclusions and future outlook

To our knowledge, we are the first to systematically detect the impact of Arp2/3 complex concentration on the structures and dynamics of actomyosin networks by using mathematical graphs and data science. These tools are shown to be useful for revealing hidden patterns in complex networks, allowing us to leverage this knowledge as crucial features to train machine learning models to forecast avalanches within actomyosin networks.

To forecast the avalanches, two types of machine learning models, the SVM and the XGBoost models, were trained under various conditions. We showed that the XGBoost model performs better at forecasting avalanches than the SVM model. However, the performance of the SVM model significantly increases when the network theory order parameters are trained in the dataset. Although the XGBoost model was sufficient compared to the SVM model in predicting avalanches in our work, in some other cases where an outstanding kernel for the SVM model was utilized, the performance of the SVM model supersedes the performance of the XGBoost model [39-41]. Therefore, ML models are not entirely a black box, when trained with physically meaningful features, they provide meaningful predictions with high probability.

Despite the difference in ML models, we have used only the features from the mechanical or topological properties of a network in forecasting avalanches with high sensitivity and specificity without any knowledge of their chemical dynamics. It is indicative that the avalanche is a mechanically dominant, common phenomenon in the simulated actomyosin systems. Although this finding is consistent with that of another work about avalanches risen from unbranched actomyosin networks [53], our independent work embraces the emergency of structural hierarchy in a network from sudden topological changes in the nanoarchitectures of branched actomyosin filaments.

**Supporting Information**

Detailed description of order parameter definitions, correlation between order parameters as well as the analysis of additional order parameters, machine learning models and simulations.

## Acknowledgement

We thank Professor Garyk Papoian from the University of Maryland for sharing the MEDYAN code with us and the comments on the manuscript. We appreciated the suggestions on running the MEDYAN code from Dr. Aravind Chandrasekaran and Dr. Haoran Ni from the Papoian group. We also thank the members from the Center for Theoretical Biological Physics (CTBP) at Rice University and its memory-focus group for sharing the stimulating ideas. Furthermore, we thank the Research Computing Data Core (RCDC) Center at the University of Houston for the computational resources. We thank the National Science Foundation for their funding support to this project (Grants No. PHY:2019745 and No. CHE:1743392). Dr. Margaret Cheung was a postdoctoral fellow from Dave's group at the University of Maryland from 2003 to 2006. Being in Dave's group was an enlightened experience in her research career. She is grateful to Dave's friendship and mentoring, and his many stimulating discussions on a wide range of topics in sciences as well as life during the lunch time at The Co-op on campus.

**TOC Graphic**

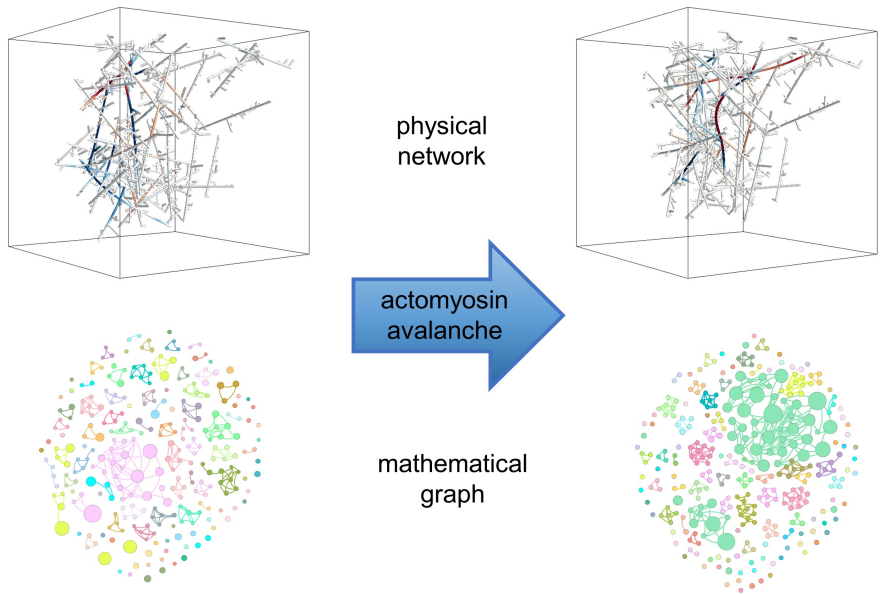